\documentstyle[prb,aps]{revtex}
\begin{document}
%
\title{Defect-Induced Resonant Tunneling of Electromagnetic Waves Through a Polariton Gap}
\author{Lev I. Deych$\dag$, A. Yamilov$\ddag$, and A.A. Lisyansky$\ddag$}
\address{$\dag$ Department of Physics, Seton Hall University, South Orange, NJ 07079\\
$\ddag$ Department of Physics, Queens College of CUNY, Flushing, NY 11367}
\maketitle
%
%

\begin{abstract}
We consider tunneling of electromagnetic waves through a polariton band gap of a 1-{\it D} chain of atoms. We analytically demonstrate that a defect embedded in the structure gives rise to the resonance transmission at the frequency of a local polariton state associated with the defect. 
\end{abstract}

\pacs{71.36+c,42.25.Bs}

The optical properties of materials with band gaps in their electromagnetic
spectrum have recently attracted a great deal of attention. It was suggested
that fundamental electromagnetic processes such as spontaneous emission,\cite
{Yablonovitch1,John1} photon-atom interaction,\cite{John1,John2} and optical
energy transfer\cite{John3} are strongly modified at band gap
frequencies. Photonic crystals, which are periodic structures with a macroscopic period,\cite
{photonic} present one of the primary examples of systems with
electromagnetic band gaps. The periodicity in photonic crystals gives rise
to allowed and forbidden bands for electromagnetic waves in basically the
same manner as periodicity in the arrangement of atoms causes the band
structure for electrons in solids.

An important property of photonic crystals is an occurrence of local photon
states with frequencies inside band gaps, when the periodic structure is
locally distorted. The fact that an isolated defect in an otherwise perfect
periodic crystal can give rise to local modes with frequencies in forbidden
gaps of a host structure is well known in solid state physics. Local
photons are similar in many aspects to other types of local states: Their
frequencies always belong to forbidden gaps; in 3-\emph{D} systems they
split off the continuous spectrum only if the ``strength'' of a defect
exceeds a certain threshold;\cite{Yablonovitch2,Joannopoulos,Figotin} and by
changing the type of the distortion one can control the position of the states
inside the gap. It is essential, however, that while all other
local states appear due to microscopic (of atomic dimensions) defects, local
photons require both a macroscopic host structure and its \emph{macroscopic%
} distortion. This fact is obviously due to the large wavelengths of
electromagnetic waves in frequency regions of interest.

Recently it was suggested that polariton gaps in regular dielectrics with
strong polar properties could generate effects similar to those attributed
to photonic crystals. \cite{Rupasov,Deych1,Podolsky,Deych2} For example,
bounded photon-atom states originally considered in Ref. [\onlinecite{John2}]
for photonic crystals were proposed in Ref. [\onlinecite{Rupasov}] for frequencies
within a polariton gap. These states are formed by an optically active impurity
atom, which possesses its own dipole moment different from that of the host
atoms. 

There exist a different type of local photon
states, that is not associated with inner optical activity of impurities. \cite{Deych1}
Such local states are analogous to
defect modes in photonic crystals, and similar to them can be used to modify spectrum of radiation of the optically active impurities. At
first glance it seems impossible, since electromagnetic waves would not
interact with microscopic objects without inner optical activity. However,
it was shown in Ref. [\onlinecite{Deych1}], that a regular {\em microscopic}defect embedded in a
crystal lattice gives rise to local states with frequencies within the
polariton gap, which are a mixture of electromagnetic component with
excitations of a crystal responsible for the polariton gap.  The most remarkable property of the local polaritons is  absence of a threshold for localization in isotropic 3-{\em D} systems, which is due to a strong van Hove singularity in the polariton density of states at the gap edge, see details in Refs. [\onlinecite{Deych1,Podolsky}].

In this  paper we show that the local states considered in Ref. [\onlinecite{Deych1}] give rise to an interesting possibility of resonance tunneling of electromagnetic waves through a polariton bandgap. 
We would like to emphasize that this tunneling proccess is remarkably different from both quantum mechanical electron tunneling\cite{electrontunneling} and photon tunneling through photonic crystals.\cite{Joannopoulos} The later proccesses can be characterized as the result of interaction between excitations and defects of comparable scales (electrons and impurities, electromagnetic waves and {\em macroscopical distortion} of photonic crystals). The tunneling studied in our paper occurs due to interaction between electromagnetic waves with {\em macroscopically} long wavelength and {\em microscopical} impurities. This process becomes possible owing to participation of the phonon component of polaritons, which mediate electromagnetic wave propagation. A similar role is played by excitons in resonance scattering of exciton-polaritons due to impurities with a short-range potential considered by Hopfield in Ref. [\onlinecite{Hopfield}].

We 
present an exact analytical solution for the transmission coefficient of a scalar wave propagating through a 1-\emph{D} chain of noninteracting atoms containing a defect. These atoms are coupled to the wave due to a dipole moment caused by their mechanical vibrations. The spectrum of the coupled excitations of the chain and the field, polaritons, have a spectral gap where the excitations can exist only in an evanescent form. 
We show, however, that a defect, embedded in such a structure, results in the resonance tunneling of waves with the transmission coefficient independent of the chain's length and being of the order of magnitude of one. 
One dimensional models usually describe tunneling processes  fairly well because, by  virtue of tunneling, the propagating wave is effectively confined in the transverse directions. In our particular situation it is also important that the local polariton states (transmitting centers) occur without a threshold in 3-\emph{D} systems as well as in 1-\emph{D} systems.\cite{Deych1,Podolsky}

 A similar model has been studied numerically in Ref. [\onlinecite{Deych2}], where a direct interaction between atoms of the chain (leading to the spatial dispersion of the chain's excitations) has been taken into account. The results of that paper suggest that though the spatial dispersion brings about some new features, it does not effect the  existence of the resonance, justifying our neglect of the inter-atomic interaction.   

The atoms in our system are represented by their vibrational polarizability $\beta _{n}$, where subindex \emph{n} represents the position of the atom in the chain. The polarizability is given by 
\begin{equation}
\beta _{n}=\frac{\alpha }{\omega ^{2}-\Omega _{n}^{2}};
\label{beta}
\end{equation}
where $\alpha $ is a coupling parameter between the dipoles and the field, and $\Omega _{n}^{2}$ represents an atom's vibrational frequency. The defect in our model differs from host atoms in this parameter only, so $\Omega _{n}^{2}=\Omega _{0}^{2}$ for all sites except of one occupied by the defect, where $\Omega _{n}^{2}=\Omega _{1}^{2}$.  
Polaritons  arise as collective excitations of dipoles
(polarization waves) coupled to the electromagnetic wave $E(x_{n})$ by means
of a coupling parameter $\alpha $. The electromagnetic subsystem is
described by the following equation of motion 
\begin{equation}
\frac{\omega ^{2}}{c^{2}}E(x)+\frac{d^{2}E}{dx^{2}}=-4\pi \frac{\omega ^{2}}{%
c^{2}}\sum_{n}P_{n}\delta (na-x),  \label{Maxwell}
\end{equation}
where the right hand side is a  polarization density caused by atomic
dipole moments, and $c$ is the speed of the wave in vacuum. The coordinate $x$ in
eq.\thinspace (\ref{Maxwell}) goes along the chain with the interatomic
distance $a$. 

We first derive an equation for the frequency of the local polariton state in the 1-\emph{D} situation. The one-dimensional nature of the model allows us to approach the problem microscopically and to take into account short-wave components of the field, including those beyond the first Brillouin band, exactly. Passing to the longwave limit at the last stage of the calculations we avoid  non-physical divergencies and renormalization procedures of the kind used in Ref. [\onlinecite{Rupasov}]. The  equation for an eigenfrequency of the local mode is
\begin{equation}
1=\Delta \Omega ^{2}\frac{a}{2\pi }\int_{-\pi /a}^{\pi /a}{\displaystyle}%
\frac{\cos (ak)-\cos (\frac{a\omega }{c})}{\left[ \omega ^{2}-\Omega
_{0}^{2}-2\Phi \cos \left( ka\right) \right] \left[ \cos (ka)-\cos (\frac{%
a\omega }{c})\right] -\frac{2\pi \alpha \omega }{c}\sin (\frac{a\omega }{c})}%
dk,
\label{eigen_freq}
\end{equation}
where $\Delta \Omega ^{2}=\Omega_1^2-\Omega_0^2$. It has a real-valued
solution only if the frequency falls into the gap between the upper and the
lower polariton branches. 
The integral in eq.(\ref{eigen_freq}) can be calculated exactly to yield
\begin{equation}
\omega ^{2}=\Omega_{1}^{2}-d^{2}\frac{\omega a}{2c}\frac{\Delta \Omega ^{2}}{\sqrt{\left(
\omega ^{2}-\Omega_{0}^{2}\right) \left( \Omega_{0}^{2}+d^{2}-\omega ^{2}\right) }},  \label{nodispersion}
\end{equation}
where we passed to the longwave limit $\omega a/c\ll 1$, and introduced parameter $d^{2}=4\pi\alpha /a$, which determines the width of the polariton gap between $\Omega _{0}^{2}$ and $\Omega _{0}^{2}+d^{2}$. The second term in eq.(\ref{nodispersion}) is small for realistic values of the parameters, therefore, the frequency of the local mode is only slightly different from the defect frequency $\Omega _{1}^{2}$. As we shall see below, this fact has a deep impact upon the transmission frequency profile of the chain. 

The field in the eigenmode corresponding to the frequency determined by eq.(\ref{nodispersion}) exponentially decreases away from the defect site:
\begin{equation}
E=E_{def}\exp\left[-\kappa a(n-n_0)\right],
\label{Edef}
\end{equation}
where $\kappa$ is an inverse localization length of the state, which in the long wavelength approximation is given by
\begin{equation}
\kappa=\frac{\omega}{c}\sqrt{\frac{\Omega_0^2+d^2-\omega^2}{\omega^2-\Omega_0^2}}.
\end{equation}

In order to consider transport properties of the model one has to  subject  eq.(\ref{Maxwell}) to the standard boundary conditions. We assume that  incident and transmitted electromagnetic waves propagate in vacuum so that the transmission, $t$, and the reflection, $r$, coefficients are defined in the usual way
\begin{equation}
E(0)=E_{in}(1+r);\quad \frac{dE}{dx}{=ik}_{\omega }E_{in}(1-r);\quad E(L)=tE_{in}\exp {(i{k}%
_{\omega }L)};\quad \frac{dE}{dx}{=i{k}_{\omega }tE_{in}\exp {(i{k}_{\omega }L),}}
\label{Eboundary}
\end{equation}
where ${k}_{\omega }=\omega /c$ is a wave number of the electromagnetic wave
in vacuum,  and $L$ is the length of the chain.
Our first goal in treating the problem of  resonance tunneling is to
convert the differential equation (\ref{Maxwell}) into the discrete form. We
can do so considering separately free propagation of electromagnetic waves
between sites and  scattering due to the interaction with a dipole moment
at the site. Let $E_{n}$ and $E_{n}^{\prime }$ be the magnitude of the
electromagnetic field and its derivative just after scattering at the $n$th
site. The electric field $E$ remains continuous at a scattering site, while
its derivative undergoes a jump, which is equal to $-4\pi {k}_{\omega
}^{2}P_{n}$. Finally, one can derive the system of difference equations
that can be written with the use of the transfer matrix, $T,$ in the form: 
\begin{equation}
v_{n+1}=T_{n}v_{n},  \label{EP}
\end{equation}
where we introduce the column vector, $v_{n},$ with components  $E_{n}$, $D_{n}$ ($D_{n}=E_{n}^{\prime }/{k}_{\omega }$). The
transfer matrix, $T_{n},$  describes the propagation of the vector
between adjacent sites: 
\begin{equation}
T _{n}=\left( 
\begin{array}{cc}
\cos k_{\omega}a & \sin k_{\omega}a \\ 
-\sin k_{\omega}a+4\pi k_{\omega}\beta _{n}\cos k_{\omega}a & \cos k_{\omega}a+4\pi k_{\omega}\beta _{n}\sin k_{\omega}a
\end{array}
\right).  \label{T reduced}
\end{equation}
The dynamical state of the system at the right end of the chain, which is
represented by the vector $v_{N}$, can be found from the initial state at
the left end, $v_{0},$ by means of the repetitive use of the transfer
matrix: $v_{N}=\prod_{1}^{N}T_{n}v_{0}$. In the case of a single impurity, the matrix product
is conveniently calculated in the basis, where the matrix $T $ for a host site is diagonal. With the use of the boundary conditions given by eq.(\ref{Eboundary}) one finally arrives at the following expression for the transmission coefficient, $t$, in the limit $\kappa L > 1$:
\begin{equation}
t=\frac{t_{0}}{\left( 1+\varepsilon \right) +i\exp \left( -ik_{\omega}L\right) \Gamma
t_{0}\cosh \left[ \kappa a(N-2n_{0}+1\right] },  \label{transmission1}
\end{equation}
where $t_0$ is the transmission coefficient in the pure system, $\Gamma =4\pi k_{\omega}\varepsilon \beta /(\sin
(k_{\omega}a)\sqrt{D})$, $D=(4\pi k_{\omega}\beta)^{2}+16\pi k_{\omega}\beta \cot (ak_{\omega})-4$,  and $\varepsilon =4\pi k_{\omega}\left( \beta _{def}-\beta
\right) /\sqrt{D}$. The last parameter reflects the difference between host atoms and the impurity, and is equal to 
\begin{equation}
\varepsilon =\frac{4\pi \alpha }{c\sqrt{D}}\omega \frac{\left( \Omega
_{1}^{2}-\Omega _{0}^{2}\right) }{\left( \omega ^{2}-\Omega _{0}^{2}\right)
\left( \omega ^{2}-\Omega _{1}^{2}\right) }  \label{epsilon}.
\end{equation}
The transmission coefficient $t_{0}$ of the pure system is given by
\begin{equation}
t_{0}=\frac{2e^{ik_{\omega}L}\exp \left( -\kappa L\right) }{1-\frac{i}{\sqrt{D}}%
\left( 2-4\pi k_{\omega}\beta \cot k_{\omega}a\right) }  \label{tpure}
\end{equation}
and exhibites a regular exponential decay. 

Eq.(\ref{transmission1}) describes the resonance tunneling of the electromagnetic
waves through the chain with the defect. The resonance occurs when 
\begin{equation}
1+\varepsilon =0,  \label{res_freq}
\end{equation}
with the transmission  becoming independent of the system size.
Substituting the definition of the parameter $\varepsilon $ from eq.(\ref
{epsilon}) in eq.(\ref{res_freq}) one arrives at the equation identical to
eq.(\ref{nodispersion}) for the frequency of the local polariton state. Typically, for  resonant tunneling, the
transmission takes a maximum value when the defect is located in the middle of
the chain, $N-2n_{0}+1=0$, and in this case 
\begin{equation}
|t_{\max }|^{2}=\frac{1}{\Gamma ^{2}}\leq 1  \label{tmax}
\end{equation}
The width of the resonance is proportional to $\Gamma t_{0}$ and
exponentially decreases with an increase of the system's size. In the
longwave limit $ak_{\omega}\ll 1$, eq.(\ref{tmax}) can be rewritten in the following
form 
\begin{equation}
|t_{max }|^{2}=1-\left( 1-2\frac{\omega _{r}^{2}-\Omega _{0}^{2}}{d^{2}}%
\right)^2,  \label{tmaxlongwave}
\end{equation}
where $\omega _{r}$ is the resonance frequency satisfying eq.(\ref{res_freq}). It is interesting to note that the transmission becomes exactly equal to
one if the resonance frequency corresponds exactly to the center of the
polariton gap. This fact has a simple physical explanation. For $\omega
_{r}^{2}=\Omega _{0}^{2}+d^{2}/2$ the inverse localization length $\kappa 
$ becomes exactly equal to the wave number $\omega _{r}/c$ of the incoming
radiation. Owing to this fact the field and its derivative inside the chain
exactly match the field and the derivative of the incoming field as though
the optical properties of the chain are identical to those in vacuum.
Consequently, the field propagates through the chain without reflection.

Equations (\ref{transmission1}) and (\ref{tmax}) demonstrate that the frequency profile of the resonance is considerably different from  a  corresponding Lorentztian profile for electronic and other known instances of tunneling. This occurs because the
parameter $\varepsilon $ diverges at $\omega =\Omega _{1}$ causing the transmission to vanish. At the
same time the resonance frequency $\omega _{r}$ is very close to $\Omega _{1}$ as it follows from eq.(\ref{nodispersion}). This results in a frequency dependence of the transmission strongly
skewed toward lower frequencies.  

Having solved the transmission problem, we can find the magnitude of the
field inside the chain in terms of the incident amplitude, $E_{in}$, at the
resonance frequency. Matching the  field in the local
polariton state given by eq.~(\ref{Edef}) with
 the outcoming field  we
have for the field amplitude at the defect atom 
\begin{equation}
E_{d}=E_{in}t\exp (-ikL)\exp \left[ (N-n_{0})\kappa a\right].
\label{defect _field}
\end{equation}
For $|t|$ being of the order of one in the resonance, this expression
describes the drastic exponential enhancement of the incident amplitude at
the defect side due to the effect of the resonance tunneling. This effect is an electromagnetic analogue of charge accumulation in the case of electron tunneling.\cite{Goldman-1}

Resonance tunneling is very sensitive to the presence of relaxation, which phenomenologically can be accounted for by adding $2i\gamma\omega$ to the denominator of the polarizability $\beta$. This will make the parameter $\epsilon$ complex valued, and the resonance condition  $Re(\varepsilon) = -1 $ may only be fulfilled if the relaxation is as small as $\gamma < (ad^2)/(4c)$. This inequality has a simple physical meaning: it ensures that the distance between the resonance frequency and $\Omega_1$, where the transmission goes to zero, is greater than the relaxation parameter $\gamma$. This is a strict condition that can only be satisfied for high frequency oscillations with large oscillator strength in molecular crystals with large molecules in an elementary cell, and respectively  large values of the interatomic spacing $a$. Another interesting opportunity can arise in so called atomic optical lattices, where atoms, trapped by a laser beam,  form a lattice with spacing practically equal to the wavelength of the trapping field.\cite{opticlattice} However, taking into account  the spatial dispersion can lead to a more favorable situation for the tunneling resonance in our model. Numerical results of Ref. [\onlinecite{Deych2}] show that spatial dispersion does not change the transmission properties significantly. Therefore, one can rely upon eq.(\ref{transmission1}) to estimate the effect of the dissipation in the presence of the spatial dispersion, assuming that it only modifies the parameter $\varepsilon$ . Eigenfrequency of the local mode in the presence of the direct inter-atomic interaction  is calculated exactly. The interaction moves the resonance frequency farther away from $\Omega_1$ undermining the influence of the damping. Adjusting  $\varepsilon$ correspondingly one can arrive at the following new condition for the resonance to survive the relaxation: $(\gamma \Omega_1)/\Phi < 1$, where $\Phi$ is the parameter of the inter-atomic interaction; it can be estimated as a bandwith of the polarization waves (in terms of squared frequencies). This condition can be  fulfilled, even for phonons with a relatively small negative spatial dispersion, and becomes even less restrictive in the case of Frenkel excitons in molecular crystals.
The imaginary part of $\varepsilon$ will prevent the exponential factor $t_0$ in eq.(\ref{transmission1}) from canceling out at the resonance. This restricts the length of the system in which the resonance can occur. The requirement that $Im(\varepsilon)$ be much smaller than $t_0$ leads to the following condition \linebreak $L\ll 1/(\kappa\mid\ln[Im(\varepsilon)]\mid)$,
 with $Im(\varepsilon)$ being again of the order of $\min[(4\gamma c)/(ad^2),(\gamma \Omega_1)/\Phi]$. 

Concluding, we showed that a regular defect atom without internal degrees of freedom and an optical activity results in resonance tunneling of electromagnetic waves through a polariton gap . Though we considered  the one-dimensional model one can expect that the existence of the effect  does not depend upon dimensionality because tunneling transport is virtually one-dimensional, and the polariton local states are present in the system of any dimension.

The authors are grateful to S. Schwarz for reading and commenting on the manuscript. This work was supported by the NSF under grant No. DMR-9632789.


\begin{thebibliography}{99}
\bibitem{Yablonovitch1}  E. Yablonovitch, Phys. Rev. Lett. {\bf 58}, 2059
(1987).

\bibitem{John1}  S. John and T. Quang, Phys. Rev. A {\bf 50}, 1764 (1994).

\bibitem{John2}  S. John and J. Wang, Phys. Rev. Lett. {\bf 64}, 2418
(1990); Phys. Rev. B {\bf 43}, 12772 (1991).

\bibitem{John3}  S. John and T. Quang, Phys. Rev. A, {\bf 52}, 4083 (1995).

\bibitem{photonic}  J.D. Joannopoulos, R.D. Meade, J.N. Winn. {\it Photonic
Crystals: Molding the Flow of Light }(Princeton Univ. Press, 1995).

\bibitem{Yablonovitch2}  E. Yablonovitch, T.J. Gmitter, R.D. Meade, A.M.
Rappe, K.D. Brommer, and J.D. Joannopoulos. Phys. Rev. Lett. {\bf 67} 3380
(1991).

\bibitem{Joannopoulos} R.D. Meade, K.D. Brommer, A.M. Rappe, and J.D.
Joannopoulos, Phys. Rev. B {\bf 44} 13772 (1991).

\bibitem{Figotin}  A. Figotin and A. Klein, J. Stat. Phys. {\bf 86}, 165
(1997).

\bibitem{Rupasov}  V.I. Rupasov and M. Singh, Phys. Rev. A, {\bf 54}, 3614
(1996); Phys. Lett. A {\bf 222}, 258 (1996).

\bibitem{Deych1}  L.I. Deych and A.A. Lisyansky, Bull. Amer. Phys. Soc. {\bf %
42}, 203 (1997); Phys. Lett. A, {\bf 240}, 329 (1998).

\bibitem{Podolsky} V.S. Podolsky, L.I. Deych, and A.A. Lisyansky, Phys.
Rev. B, {\bf 57}, 5168 (1998).

\bibitem{Deych2} L.I. Deych and A.A. Lisyansky, Phys. Lett. A, {\bf 243},
156 (1998).

\bibitem{electrontunneling} A.V. Chaplik and M.V. Entin, Zh. Eksp. \& Teor.
Fiz. {\bf 67}, 208 (1974) [Sov. Phys.- JETP {\bf 40}]; I.M. Lifshitz and V. Ya Kirpichenkov, Zh. Eksp. \& Teor. Fiz. {\bf 77}, 989 (1979) [Sov. Phys.- JETP {\bf 50}].

\bibitem{IEM}  R. Bellman and G. Wing, {\it An introduction to invariant
embedding} (Wiley, New York, 1976).

\bibitem{Hopfield} J.J. Hopfield, Phys. Rev., {\bf 182}, 945 (1969).

\bibitem{Goldman-1} V.J. Goldman, D.C. Tsui, and J.E. Cunningham, Phys.
Rev. Lett. {\bf 58}, 1256 (1987); V.J. Goldman, D.C. Tsui, and J.E. Cunningham, Phys.
Rev. B. {\bf 35}, 9387 (1987); A. Zaslavsky, V.J. Goldman, D.C. Tsui, and J.E.
Cunningham, Appl. Phys. Lett. {\bf 53}, 1408 (1988).

\bibitem{opticlattice} M. Weidem\"uller, A. Hemmerich, A. G\"orlitz, T. Esslinger, and T.W. H\"ansch, Phys. Rev. Lett., {\bf 75}, 4583 (1995); G. Birkl, M. Gatzke, I.H. Deutsch, S.L. Rolston, and W.D. Phillips, Phys. Rev. Lett., {\ bf 75}, 2823, (1995); J.N. Tan, J.J. Bollinger, B. Jelenkovic, and D.J. Wineland, Phys. Rev. Lett., {\bf 75}, 4198 (1995).

\end{thebibliography}
\end{document}